\input harvmac
\input epsf
\input amssym
%
\noblackbox
\newcount\figno
\figno=0
\def\fig#1#2#3{
\par\begingroup\parindent=0pt\leftskip=1cm\rightskip=1cm\parindent=0pt
\baselineskip=11pt
\global\advance\figno by 1
\midinsert
\epsfxsize=#3
\centerline{\epsfbox{#2}}
\vskip -21pt
{\bf Fig.\ \the\figno: } #1\par
\endinsert\endgroup\par
}
\def\figlabel#1{\xdef#1{\the\figno}}
\def\encadremath#1{\vbox{\hrule\hbox{\vrule\kern8pt\vbox{\kern8pt
\hbox{$\displaystyle #1$}\kern8pt}
\kern8pt\vrule}\hrule}}

\def\frac#1#2{{#1 \over #2}}

\def\semi{\subset\kern-1em\times\;}
\def\bar#1{\overline{#1}}
\def\sqr#1#2{{\vcenter{\vbox{\hrule height.#2pt
\hbox{\vrule width.#2pt height#1pt \kern#1pt \vrule width.#2pt}
\hrule height.#2pt}}}}

               \def\CQ{{\cal Q}}

\def\ra{{\rightarrow }}

\def\ad{\bar a}

\def\ts{{\tilde{s}}}

\def\Qb{\overline{Q}}

\def\psit{\tilde{\psi}}
\def\phit{\tilde{\phi}}

\def\thetat{\tilde{\theta}}

\def\thetah{\hat{\theta}}
\def\phih{\hat{\phi}}

\def\tt{\tilde{t}}
\def\rt{\tilde{r}}

\def\Rt{\tilde{R}}

\def\tZ{ \tilde{Z}}

%

\def\Qb{\overline{Q}}

\def\psit{\tilde{\psi}}
\def\phit{\tilde{\phi}}

\def\coeff#1#2{\relax{\textstyle {#1 \over #2}}\displaystyle}
\def\ZZ{\Bbb{Z}}
\def\IC{\Bbb{C}}
\def\ID{\Bbb{D}}
\def\IF{\Bbb{F}}
\def\IH{\Bbb{H}}
\def\II{\Bbb{I}}
\def\IN{\Bbb{N}}
\def\IP{\Bbb{P}}
\def\IQ{\Bbb{Q}}
\def\IR{\Bbb{R}}

\def\Sigmat{\tilde{\Sigma}}

\def\bfone{\relax{\rm 1\kern-.35em 1}}
\def\IC{\relax\,\hbox{$\inbar\kern-.3em{\rm C}$}}
\def\ID{\relax{\rm I\kern-.18em D}}
\def\IF{\relax{\rm I\kern-.18em F}}
\def\IH{\relax{\rm I\kern-.18em H}}
\def\II{\relax{\rm I\kern-.17em I}}
\def\IN{\relax{\rm I\kern-.18em N}}
\def\IP{\relax{\rm I\kern-.18em P}}
\def\IQ{\relax\,\hbox{$\inbar\kern-.3em{\rm Q}$}}

\def\IR{\relax{\rm I\kern-.18em R}}
\font\cmss=cmss10 \font\cmsss=cmss10 at 7pt
\def\ZZ{\relax\ifmmode\mathchoice
{\hbox{\cmss Z\kern-.4em Z}}{\hbox{\cmss Z\kern-.4em Z}}
{\lower.9pt\hbox{\cmsss Z\kern-.4em Z}}
{\lower1.2pt\hbox{\cmsss Z\kern-.4em Z}}\else{\cmss Z\kern-.4em
Z}\fi}

%

\lref\AdSBH{
J.~B.~Gutowski and H.~S.~Reall,
  ``General supersymmetric AdS(5) black holes,''
  JHEP {\bf 0404}, 048 (2004)
  [arXiv:hep-th/0401129].
}
\lref\myers{ R.~C.~Myers,
``Dielectric-branes,''
JHEP {\bf 9912}, 022 (1999)
[arXiv:hep-th/9910053].
 }
\lref\supertube{
D.~Mateos and P.~K.~Townsend,
``Supertubes,''
Phys.\ Rev.\ Lett.\  {\bf 87}, 011602 (2001)
[arXiv:hep-th/0103030].
}

\lref\mathurlunin{
O.~Lunin and S.~D.~Mathur,
``Metric of the multiply wound rotating string,''
Nucl.\ Phys.\ B {\bf 610}, 49 (2001)
[arXiv:hep-th/0105136].
}

\lref\LuninJY{
O.~Lunin and S.~D.~Mathur,
``AdS/CFT duality and the black hole information paradox,''
Nucl.\ Phys.\ B {\bf 623}, 342 (2002)
[arXiv:hep-th/0109154].
}

\lref\mathurstretch{
O.~Lunin and S.~D.~Mathur,
 ``Statistical interpretation of Bekenstein entropy for systems with a
stretched horizon,''
Phys.\ Rev.\ Lett.\  {\bf 88}, 211303 (2002)
[arXiv:hep-th/0202072].
}

\lref\LuninBJ{
O.~Lunin, S.~D.~Mathur and A.~Saxena,
``What is the gravity dual of a chiral primary?,''
Nucl.\ Phys.\ B {\bf 655}, 185 (2003)
[arXiv:hep-th/0211292].
}

\lref\lmm{
  O.~Lunin, J.~Maldacena and L.~Maoz,
  ``Gravity solutions for the D1-D5 system with angular momentum,''
  arXiv:hep-th/0212210.
}

\lref\mathur{
S.~D.~Mathur, A.~Saxena and Y.~K.~Srivastava,
``Constructing 'hair' for the three charge hole,''
arXiv:hep-th/0311092.
}
\lref\MathurSV{
S.~D.~Mathur,
``Where are the states of a black hole?,''
arXiv:hep-th/0401115.
}

\lref\bmpv{
J.~C.~Breckenridge, R.~C.~Myers, A.~W.~Peet and C.~Vafa,
``D-branes and spinning black holes,''
Phys.\ Lett.\ B {\bf 391}, 93 (1997)
[arXiv:hep-th/9602065].
}

\lref\emparan{
R.~Emparan, D.~Mateos and P.~K.~Townsend,
``Supergravity supertubes,''
JHEP {\bf 0107}, 011 (2001)
[arXiv:hep-th/0106012].
}

\lref\MateosPR{
D.~Mateos, S.~Ng and P.~K.~Townsend,
``Tachyons, supertubes and brane/anti-brane systems,''
JHEP {\bf 0203}, 016 (2002)
[arXiv:hep-th/0112054].
}

\lref\Buscher{T.~H.~Buscher, Phys.\ Lett.\ B {\bf 159}, 127
(1985),\ B {\bf 194}, 59 (1987),\ B  {\bf 201}, 466 (1988),
}

\lref\MeessenQM{ P.~Meessen and T.~Ortin, ``An Sl(2,Z) multiplet
of nine-dimensional type II supergravity theories,'' Nucl.\ Phys.\
B {\bf 541}, 195 (1999) [arXiv:hep-th/9806120].
}

\lref\CallanHN{ C.~G.~Callan, J.~M.~Maldacena and A.~W.~Peet,
``Extremal Black Holes As Fundamental Strings,'' Nucl.\ Phys.\ B
{\bf 475}, 645 (1996) [arXiv:hep-th/9510134].
}

\lref\DabholkarNC{ A.~Dabholkar, J.~P.~Gauntlett, J.~A.~Harvey and
D.~Waldram, ``Strings as Solitons \& Black Holes as Strings,''
Nucl.\ Phys.\ B {\bf 474}, 85 (1996) [arXiv:hep-th/9511053].
}

\lref\CveticXH{ M.~Cvetic and F.~Larsen, ``Near horizon geometry
of  rotating black holes in five dimensions,'' Nucl.\ Phys.\ B
{\bf 531}, 239 (1998) [arXiv:hep-th/9805097].
}

\lref\EmparanWN{
R.~Emparan and H.~S.~Reall,
``A rotating black ring in five dimensions,''
Phys.\ Rev.\ Lett.\  {\bf 88}, 101101 (2002)
[arXiv:hep-th/0110260].
}

\lref\ReallBH{
H.~S.~Reall,
``Higher dimensional black holes and supersymmetry,''
Phys.\ Rev.\ D {\bf 68}, 024024 (2003)
[arXiv:hep-th/0211290].
}

\lref\ElvangMJ{
H.~Elvang and R.~Emparan,
 ``Black rings, supertubes, and a stringy resolution of black hole
non-uniqueness,''
JHEP {\bf 0311}, 035 (2003)
[arXiv:hep-th/0310008].
}

\lref\EmparanWY{
R.~Emparan,
``Rotating circular strings, and infinite non-uniqueness of black rings,''
arXiv:hep-th/0402149.
}

\lref\BenaWV{
  I.~Bena,
  ``Splitting hairs of the three charge black hole,''
  Phys.\ Rev.\ D {\bf 70}, 105018 (2004)
  [arXiv:hep-th/0404073].
}

\lref\BalasubramanianRT{ V.~Balasubramanian, J.~de Boer,
E.~Keski-Vakkuri  and S.~F.~Ross, ``Supersymmetric conical
defects: Towards a string theoretic description  of
black hole formation,''
Phys.\ Rev.\ D {\bf 64}, 064011 (2001) [arXiv:hep-th/0011217].
}

\lref\MaldacenaDR{ J.~M.~Maldacena and L.~Maoz,
``De-singularization  by rotation,'' JHEP {\bf 0212}, 055 (2002)
[arXiv:hep-th/0012025].
}

\lref\BW{ I.~Bena and N.~P.~Warner, ``One Ring to Rule Them All
... and  in the Darkness Bind Them?,'' arXiv:hep-th/0408106.
}

\lref\MaharanaMY{
  J.~Maharana and J.~H.~Schwarz,
  ``Noncompact symmetries in string theory,''
  Nucl.\ Phys.\ B {\bf 390}, 3 (1993)
  [arXiv:hep-th/9207016].
}

\lref\GauntlettQY{
  J.~P.~Gauntlett and J.~B.~Gutowski,
  ``General concentric black rings,''
  Phys.\ Rev.\ D {\bf 71}, 045002 (2005)
  [arXiv:hep-th/0408122].
}

\lref\ElvangYY{ H.~Elvang, ``A charged rotating black ring,''
Phys.\ Rev. \ D {\bf 68}, 124016 (2003) [arXiv:hep-th/0305247].
}

\lref\StromingerSH{ A.~Strominger and C.~Vafa, ``Microscopic
Origin  of the Bekenstein-Hawking Entropy,'' Phys.\ Lett.\ B {\bf
379}, 99 (1996) [arXiv:hep-th/9601029].
}

\lref\ConstableDJ{ N.~R.~Constable, C.~V.~Johnson and R.~C.~Myers,
``Fractional branes and the entropy of 4D black holes,'' JHEP {\bf
0009}, 039 (2000) [arXiv:hep-th/0008226].
}

\lref\JohnsonGA{ C.~V.~Johnson, R.~R.~Khuri and R.~C.~Myers,
``Entropy of 4D Extremal Black Holes,''
Phys.\ Lett.\ B {\bf 378}, 78 (1996) [arXiv:hep-th/9603061].
}

\lref\MaldacenaGB{ J.~M.~Maldacena and A.~Strominger,
``Statistical  Entropy of Four-Dimensional Extremal Black Holes,''
Phys.\ Rev.\ Lett.\  {\bf 77}, 428 (1996) [arXiv:hep-th/9603060].
}

\lref\CveticXH{ M.~Cvetic and F.~Larsen, ``Near horizon geometry
of  rotating black holes in five dimensions,'' Nucl.\ Phys.\ B
{\bf 531}, 239 (1998) [arXiv:hep-th/9805097].
}

\lref\GiustoID{ S.~Giusto, S.~D.~Mathur and A.~Saxena, ``Dual
geometries for a set of 3-charge microstates,''
arXiv:hep-th/0405017.
}

\lref\GiustoIP{ S.~Giusto, S.~D.~Mathur and A.~Saxena, ``3-charge
geometries and their CFT duals,'' arXiv:hep-th/0406103.
}

\lref\LarsenUK{ F.~Larsen and E.~J.~Martinec, ``U(1) charges and
moduli in the D1-D5 system,'' JHEP {\bf 9906}, 019 (1999)
[arXiv:hep-th/9905064].
}

\lref\SeibergXZ{ N.~Seiberg and E.~Witten, ``The D1/D5 system and
singular CFT,'' JHEP {\bf 9904}, 017 (1999)
[arXiv:hep-th/9903224].
}

\lref\LuninUU{ O.~Lunin, ``Adding momentum to D1-D5 system,'' JHEP
{\bf 0404}, 054 (2004) [arXiv:hep-th/0404006].
}

\lref\KalloshUY{ R.~Kallosh and B.~Kol, ``E(7) Symmetric Area of
the Black Hole Horizon,'' Phys.\ Rev.\ D {\bf 53}, 5344 (1996)
[arXiv:hep-th/9602014].
}

\lref\BertoliniYA{ M.~Bertolini and M.~Trigiante, ``Microscopic
entropy of the most general four-dimensional BPS black  hole,''
JHEP {\bf 0010}, 002 (2000) [arXiv:hep-th/0008201].
}
\lref\BertoliniEI{ M.~Bertolini and M.~Trigiante, ``Regular BPS
black  holes: Macroscopic and microscopic description of the
generating solution,''
Nucl.\ Phys.\ B {\bf 582}, 393 (2000) [arXiv:hep-th/0002191].
}
\lref\MaldacenaDE{ J.~M.~Maldacena, A.~Strominger and E.~Witten,
``Black hole entropy in M-theory,'' JHEP {\bf 9712}, 002 (1997)
[arXiv:hep-th/9711053].
}

\lref\DavidWN{ J.~R.~David, G.~Mandal and S.~R.~Wadia,
``Microscopic  formulation of black holes in string theory,''
Phys.\ Rept.\ {\bf 369}, 549 (2002) [arXiv:hep-th/0203048].
}

\lref\AharonyTI{ O.~Aharony, S.~S.~Gubser, J.~M.~Maldacena,
H.~Ooguri and  Y.~Oz, ``Large N field theories, string theory and
gravity,'' Phys.\ Rept.\  {\bf 323}, 183 (2000)
[arXiv:hep-th/9905111].
}

\lref\BanadosWN{ M.~Banados, C.~Teitelboim and J.~Zanelli, ``The
Black hole  in three-dimensional space-time,'' Phys.\ Rev.\ Lett.\
{\bf 69}, 1849 (1992) [arXiv:hep-th/9204099].
}

\lref\BrownNW{ J.~D.~Brown and M.~Henneaux, ``Central Charges In
The  Canonical Realization Of Asymptotic Symmetries: An
Example From Three-Dimensional Gravity,''
Commun.\ Math.\ Phys.\  {\bf 104}, 207 (1986).
}

\lref\KutasovZH{ D.~Kutasov, F.~Larsen and R.~G.~Leigh, ``String
theory in magnetic  monopole backgrounds,'' Nucl.\ Phys.\ B {\bf
550}, 183 (1999) [arXiv:hep-th/9812027].
}

\lref\LarsenDH{ F.~Larsen and E.~J.~Martinec, ``Currents and
moduli in the (4,0)  theory,'' JHEP {\bf 9911}, 002 (1999)
[arXiv:hep-th/9909088].
}

\lref\min{ J.~P.~Gauntlett, J.~B.~Gutowski, C.~M.~Hull,
S.~Pakis and H.~S.~Reall,
``All supersymmetric solutions of minimal supergravity in five dimensions,''
Class.\ Quant.\ Grav.\  {\bf 20}, 4587 (2003)
[arXiv:hep-th/0209114].
}

\lref\GutowskiRG{
J.~B.~Gutowski, D.~Martelli and H.~S.~Reall,
``All supersymmetric solutions of minimal supergravity in six
dimensions,'' Class.\ Quant.\ Grav.\  {\bf 20}, 5049 (2003)
[arXiv:hep-th/0306235].
}

\lref\usc{
C.~N.~Gowdigere, D.~Nemeschansky and N.~P.~Warner,
``Supersymmetric solutions with fluxes from algebraic Killing spinors,''
arXiv:hep-th/0306097.
}

\lref\MathurZP{
  S.~D.~Mathur,
  ``The fuzzball proposal for black holes: An elementary review,''
  arXiv:hep-th/0502050.
}

\lref\GiustoKJ{
  S.~Giusto and S.~D.~Mathur,
  ``Geometry of D1-D5-P bound states,''
  arXiv:hep-th/0409067.
}

\lref\GiustoIP{
  S.~Giusto, S.~D.~Mathur and A.~Saxena,
  ``3-charge geometries and their CFT duals,''
  Nucl.\ Phys.\ B {\bf 710}, 425 (2005)
  [arXiv:hep-th/0406103].
}

\lref\GiustoID{
  S.~Giusto, S.~D.~Mathur and A.~Saxena,
  ``Dual geometries for a set of 3-charge microstates,''
  Nucl.\ Phys.\ B {\bf 701}, 357 (2004)
  [arXiv:hep-th/0405017].
}

\lref\LuninUU{
  O.~Lunin,
  ``Adding momentum to D1-D5 system,''
  JHEP {\bf 0404}, 054 (2004)
  [arXiv:hep-th/0404006].
}

\lref\BenaTK{
  I.~Bena and P.~Kraus,
  ``Microscopic description of black rings in AdS/CFT,''
  JHEP {\bf 0412}, 070 (2004)
  [arXiv:hep-th/0408186].
}

\lref\CyrierHJ{
  M.~Cyrier, M.~Guica, D.~Mateos and A.~Strominger,
  ``Microscopic entropy of the black ring,''
  arXiv:hep-th/0411187.
}

\lref\BenaWT{
  I.~Bena and P.~Kraus,
  ``Three charge supertubes and black hole hair,''
  Phys.\ Rev.\ D {\bf 70}, 046003 (2004)
  [arXiv:hep-th/0402144].
}

\lref\ElvangDS{
  H.~Elvang, R.~Emparan, D.~Mateos and H.~S.~Reall,
  ``Supersymmetric black rings and three-charge supertubes,''
  Phys.\ Rev.\ D {\bf 71}, 024033 (2005)
  [arXiv:hep-th/0408120].
}

\lref\ElvangRT{
  H.~Elvang, R.~Emparan, D.~Mateos and H.~S.~Reall,
  ``A supersymmetric black ring,''
  Phys.\ Rev.\ Lett.\  {\bf 93}, 211302 (2004)
  [arXiv:hep-th/0407065].
}

\lref\BenaTD{
  I.~Bena, C.~W.~Wang and N.~P.~Warner,
  ``Black rings with varying charge density,''
  arXiv:hep-th/0411072.
}

\lref\BenaDE{
  I.~Bena and N.~P.~Warner,
  ``One ring to rule them all ... and in the darkness bind them?,''
  arXiv:hep-th/0408106.
}

\lref\GrossHB{
  D.~J.~Gross and M.~J.~Perry,
  ``Magnetic Monopoles In Kaluza-Klein Theories,''
  Nucl.\ Phys.\ B {\bf 226}, 29 (1983).
}

\lref\SorkinNS{
  R.~d.~Sorkin,
  ``Kaluza-Klein Monopole,''
  Phys.\ Rev.\ Lett.\  {\bf 51}, 87 (1983).
}

\lref\KS{
I.~R.~Klebanov and M.~J.~Strassler,
``Supergravity and a confining gauge theory: Duality cascades and
$\chi$SB-resolution of naked singularities,''
JHEP {\bf 0008}, 052 (2000)
[arXiv:hep-th/0007191].

I.~R.~Klebanov and A.~A.~Tseytlin,
``Gravity duals of supersymmetric SU(N) x SU(N+M) gauge theories,''
Nucl.\ Phys.\ B {\bf 578}, 123 (2000)
[arXiv:hep-th/0002159].
}

\lref\GiustoXM{
  S.~Giusto and S.~D.~Mathur,
  ``Fuzzball geometries and higher derivative corrections for extremal holes,''
  arXiv:hep-th/0412133.
}

\lref\denef{
  F.~Denef,
   ``Supergravity flows and D-brane stability,''
  JHEP {\bf 0008}, 050 (2000)
  [arXiv:hep-th/0005049].

  B.~Bates and F.~Denef,
   ``Exact solutions for supersymmetric stationary black hole composites,''
  arXiv:hep-th/0304094.

  F.~Denef,
   ``Quantum quivers and Hall/hole halos,''
  JHEP {\bf 0210}, 023 (2002)
  [arXiv:hep-th/0206072].
}

\lref\tong{
  D.~Tong,
``NS5-branes, T-duality and worldsheet instantons,''
  JHEP {\bf 0207}, 013 (2002)
  [arXiv:hep-th/0204186].
}
\lref\GaiottoGF{
  D.~Gaiotto, A.~Strominger and X.~Yin,
  ``New connections between 4D and 5D black holes,''
  arXiv:hep-th/0503217.
}

\lref\GauntlettQY{
  J.~P.~Gauntlett and J.~B.~Gutowski,
``General concentric black rings,''
  Phys.\ Rev.\ D {\bf 71}, 045002 (2005)
  [arXiv:hep-th/0408122].
}

\lref\GauntlettWH{
  J.~P.~Gauntlett and J.~B.~Gutowski,
  ``Concentric black rings,''
  Phys.\ Rev.\ D {\bf 71}, 025013 (2005)
  [arXiv:hep-th/0408010].
}

\lref\BenaAY{
  I.~Bena and P.~Kraus,
  ``Microstates of the D1-D5-KK system,''
  arXiv:hep-th/0503053.
}

\lref\KrausGH{
  P.~Kraus and F.~Larsen,
  ``Attractors and black rings,''
  arXiv:hep-th/0503219.
}

\lref\HorowitzJE{
  G.~T.~Horowitz and H.~S.~Reall,
  ``How hairy can a black ring be?,''
  Class.\ Quant.\ Grav.\  {\bf 22}, 1289 (2005)
  [arXiv:hep-th/0411268].
}

\lref\GaiottoXT{
  D.~Gaiotto, A.~Strominger and X.~Yin,
  ``5D Black Rings and 4D Black Holes,''
  arXiv:hep-th/0504126.
}

\lref\ElvangSA{
  H.~Elvang, R.~Emparan, D.~Mateos and H.~S.~Reall,
  ``Supersymmetric 4D rotating black holes from 5D black rings,''
  arXiv:hep-th/0504125.
}


\Title{
  \vbox{\baselineskip12pt \hbox{hep-th/0504142}
  \hbox{UCLA-05-TEP-13}
  \vskip-.5in}
}
{\vbox{\vskip -1.0cm
\centerline{\hbox{Black Rings in Taub-NUT}}}}
\vskip -.3cm
\centerline{Iosif~Bena${}^{(1)}$, Per~Kraus${}^{(1)}$, and
Nicholas P.\ Warner${}^{(2)}$}
\bigskip
\centerline{{${}^{(1)}$\it Department of Physics and Astronomy}}
\centerline{{\it University of California}}
\centerline{{\it Los Angeles, CA  90095, USA}}
\medskip
\centerline{{${}^{(2)}$\it Department of Physics and Astronomy}}
\centerline{{\it University of Southern California}}
\centerline{{\it Los Angeles, CA 90089-0484, USA}}
\medskip

\bigskip
\bigskip

\baselineskip14pt
\noindent

We construct the most generic three-charge, three-dipole-charge, BPS
black-ring solutions in a Taub-NUT background. These solutions depend 
on seven charges
and six moduli, 
and interpolate between a four-dimensional black hole and a five-dimensional
black ring. They are also instrumental in determining the correct microscopic
description of the five-dimensional BPS black rings.

\vskip .3in


\vfill\eject

\baselineskip 13pt

\newsec{Introduction}

A new window into black-hole physics has opened up recently with
the prediction \refs{\BenaWT,\BenaWV} and subsequent discovery 
\refs{\ElvangRT,\BW,\ElvangDS,\GauntlettQY} of supergravity  solutions for 
BPS black rings and three charge supertubes.
Among other things, these solutions provide
counterexamples to black-hole uniqueness, allow for a more
detailed map between bulk and boundary states in the AdS/CFT
correspondence, and lead to a generalized version of the black
hole attractor mechanism \KrausGH.  Undoubtedly, much more remains to be
discovered.

Our focus here is on presenting the details of a solution that two
of the present authors conjectured in an earlier paper \BenaAY.
That paper gave a solution representing a two-charge supertube
wrapped around the fibre of a Taub-NUT space.  In the type IIB
frame these solutions represent microstates of the D1-D5-KK
system, and it was shown that the solutions are non-singular and
in perfect accord with CFT expectations.  Since the size of the
Taub-NUT circle stabilizes at large radius, this solution
is an asymptotically flat four dimensional solution
carrying angular momentum.    In the conclusion of \BenaAY\ it was
noted that the natural extension is to replace the two-charge
supertube by a three-charge black ring, and that the corresponding
solution would represent a four-dimensional BPS black hole
carrying angular momentum. Such solutions have also been explored
in \denef.

Here we will present the  general version of this class of
solutions. It turns out to be quite straightforward to construct,
either by solving directly the equations that give
five-dimensional BPS three-charge solutions \refs{\AdSBH,\BW}, or
by using the general tools for finding BPS solutions with
Gibbons-Hawking base spaces \GauntlettQY.  In particular, these
solutions are specified by eight harmonic functions that can be
chosen freely, subject to the constraints of smoothness and
asymptotic flatness. The coefficients in the harmonic functions
represent combinations of charges and moduli, whose
interpretations we spell out. We also point out that the absence
of closed timelike curves is governed by two functions, one of
which is the $E_{7(7)}$ quartic invariant of the eight harmonic
functions that specify the solution.

One of our aims is to resolve some of the confusion regarding the
charges and the microscopic interpretation of black rings.  Based on the 
D-brane physics underlying the existence of three charge supertubes 
and black rings \BenaWT, in 
\BW\ it was argued that the charge of BPS black-ring solutions has
a component corresponding to charge dissolved into fluxes, and
another component that comes from the local charge of the  ring.
In \BenaTK\  it was then argued that one can give a microscopic
description of the black ring entropy by realizing that the
near-ring geometry is similar to the five-dimensional lift of a
four-dimensional black hole with charges given by the local
charges of the ring. The $E_{7(7)}$ invariant microscopic entropy
derived from the CFT of this four-dimensional black hole correctly
reproduces the black ring entropy.

However, these local charges differ from the black-ring charges
measured at infinity, and in \CyrierHJ\ it was subsequently
proposed to base the CFT description of black rings on these
asymptotic charges. Taking into account the zero point shift of
the level number,  these authors also reproduced the black ring
entropy formula.  The relation of these two approaches was, and
remains, unclear. Moreover, in \HorowitzJE\ it was argued that
using a  certain
 mathematically correct definition of charge, one measures the same charge both
at infinity and on the ring. This was then used to argue that
the charges of the black ring are not the local charges introduced
in \BW, but the charges measured at infinity,
and that black ring solutions have no charge dissolved in fluxes.

Knowing the solution for a black ring in Taub-NUT, as proposed in
\BenaAY, will allow us to give a much cleaner derivation of the
relation between the four-dimensional black hole and the
five-dimensional black ring\foot{Note that this type of approach
was exploited recently in \GaiottoGF\  to relate the
five-dimensional BMPV black hole to four-dimensional black holes,
and to thus  give a five-dimensional interpretation to the
topological string partition function.}. By adjusting moduli we
can continuously tune the radius of the ring. For small radius the
ring sits near the locally flat origin of Taub-NUT, and so reduces
to a five-dimensional black ring.  For large radius the ring sits
in the region where the Taub-NUT circle stabilizes at fixed size,
and so the solution becomes a four-dimensional black hole.  The
entropy depends only upon quantized charges, and so it is constant
during the interpolation from small to large radius. One can then
check in the large radius limit whether the microscopic charges of
the ring, and its four-dimensional black-hole description,
correspond to the local charges used in \BW\ and \BenaTK\ or to the
asymptotic charges used in  \CyrierHJ. We will see that it is the
former, and we take this as evidence for the correctness of the
logic of \BenaTK.   It remains an interesting question to explain
the success of the alternative computation, since it seems
unlikely to be an accident.

This interpolation also establishes that the microscopic charges
of the five-dimensional BPS black ring are the local charges
discussed in \BW, and not the charges measured at infinity. Hence,
the charge introduced in \HorowitzJE,  while mathematically
well-defined, does not measure the microscopic  charge of the
black rings, but some other quantity. The meaning of this quantity
is clear when the black ring is in the near-tip region: it is the
charge measured asymptotically in  five dimensions. However, in
the limit when the black ring is far away from the tip, in the
four-dimensional region, the meaning of this quantity is much less
clear. 

The remainder of this paper is organized as follows. In section 2
we focus on explaining the black-ring entropy by adjusting the
ring radius.  We will do this in the context of the simplest
version of the black ring in Taub-NUT, so as not to obscure the
basic physics.  In section 3 we present the form of the general
three-charge solutions with a Gibbons-Hawking base. We re-derive
the equations and form of the solution, and give a systematic
method for solving these equations. In section 4 we use this to
construct the general black ring in a Taub-NUT background.  This
solution contains a number of adjustable moduli, whose physical
meaning we discuss.

\noindent {\it Note:} As this work was being completed, two papers
\refs{\ElvangSA,\GaiottoXT} appeared which also
discuss the black ring in Taub-NUT.

\newsec{Black ring in Taub-NUT}

In this section we will give a physically motivated derivation of
the simplest black ring in Taub-NUT, and  show how adjusting the
radius of the black ring allows one to interpolate between a black
ring in five dimensions and a black hole in four dimensions.

Our starting point is the five-dimensional black ring presented in
\refs{\ElvangRT,\BW,\ElvangDS,\GauntlettQY}.
In this section we will only pay attention to the metric, which
captures the basic physics of the problem, and defer discussion of
the field strengths and moduli to the next section.  The black
ring metric is:
\eqn\za{\eqalign{ ds^2 &= -(Z_1 Z_2 Z_3)^{-2/3}(dt+k)^2+ (Z_1 Z_2
Z_3)^{1/3} \left( d\rt^2 + \rt^2(d\thetat^2 + \sin^2 \thetat
d\psit^2 +\cos^2 \thetat d\phit^2)\right) \cr Z_I & = 1 + {\Qb_I
\over \Sigmat} + \half C_{IJK} q^J q^K {\rt^2 \over \Sigmat^2} \cr
k & = -{\rt^2 \over 2 \Sigmat^2} \left(q^I \Qb_I + {2 q^1 q^2 q^3
\rt^2 \over \Sigmat}\right)(\cos^2 \thetat \, d\phit + \sin^2
\thetat \, d\psit) -3  J_T \, {2 \rt^2 \sin^2 \thetat\over \Sigmat
(\rt^2 +\Rt^2 +\Sigmat)}\, d\psit \,,}}
where $C_{IJK}=1$ for $(IJK)=(123)$ and permutations thereof,
\eqn\zb{ \Sigmat = \sqrt{(\rt^2-\Rt^2)^2 + 4 \Rt^2 \rt^2 \cos^2
\thetat }\,,}
and $J_T$ is the difference between the two angular momenta of the
ring: $J_T \equiv J_{\psit} -J_{\phit}~$. The radius of the ring
is $\Rt$, and is related to $J_T$ by
\eqn\jtt{J_T = (q^1 +q^2 +q^3) \Rt^2 .}
We wrote the solution in
terms of the ring charges $\Qb_I$.  As we have noted, for the 5D
black ring these differ from the charges measured at infinity,
which are $Q_I = \Qb_I + \half C_{IJK}q^J q^K$.

It is convenient to choose  units such that $G_5 = {\pi \over 4}$, and
choose the three $T^2$'s that appear in the M-theory lift of this
solution to have equal size.
In these units, the charges $Q_I$, $\Qb_I$ and $q^I$ are quantized as
integers.

We now perform a change of coordinates, to bring the black ring to
a form in which we can easily include it in Taub-NUT.  We define
\eqn\eb{\eqalign{ \phi & = \phit-\psit ,\quad \psi =2\psit , \quad
\theta = 2\thetat, \quad \rho = {\rt^2 \over 4 }~.}}
The coordinate ranges are given by
\eqn\ec{\eqalign{
\theta &  \in(0,\pi), \quad (\psi,\phi) \cong (\psi +4\pi,\phi)
\cong( \psi, \phi+2\pi)~.}}
In the new coordinates the black-ring metric is
\eqn\zd{\eqalign{ ds^2 &= -(Z_1 Z_2 Z_3)^{-2/3}(dt+k)^2+ (Z_1 Z_2
Z_3)^{1/3} h_{mn}dx^m dx^n  \,, \cr Z_I & = 1 + {\Qb_I \over
4\Sigma} + \half C_{IJK} q^J q^K {\rho \over 4\Sigma^2} \,, \cr
k&= \mu\left(d\psi+(1+\cos \theta) d\phi \right)+\omega  \,, \cr \mu & = -{1 \over
16}{\rho \over \Sigma^2} \left(q^I\Qb_I + {2 q^1 q^2 q^3 \rho
\over \Sigma}\right) +
{J_T \over 16 R}
\left(1-{\rho\over
\Sigma}-{R \over \Sigma}\right) \,,
\cr
\omega & =-
 {J_T \rho \over 4 \Sigma (\rho +R + \Sigma)}
\sin^2\theta d\phi\,,}}
with
\eqn\ze{\eqalign{h_{mn}dx^m dx^n &= V^{-1}\left(d\psi+ (1+\cos \theta)
d\phi\right)^2 +V(d\rho^2 + \rho^2(d\theta^2 +\sin^2 \theta d\phi^2))\cr
 V &={1\over \rho}, \quad  \Sigma = \sqrt{\rho^2 +R^2 + 2R \rho\cos\theta},
\quad R = {\Rt^2 \over 4}~.}}

The first line of \ze\ is just flat space written in
Gibbons-Hawking coordinates.  In these coordinates, the ring is
sitting at a distance $R$ along the negative $z$ axis of the
three-dimensional base.
To change the four-dimensional base metric into Taub-NUT one needs
to add a $1$ to the harmonic function $V$. This is easily
accomplished using the general results of Gauntlett and Gutowski
on solutions with Gibbons-Hawking base \GauntlettQY, which we also
re-derive in the next section.

A general supersymmetric solution can be written in terms of
harmonic functions $K^I,~L_I,~M$ and $V$ as (our conventions
differ from \GauntlettQY):
\eqn\zf{\eqalign{ Z_I &= {1 \over 2}H^{-1} C_{IPQ}K^P K^Q +L_I \,, \cr
\mu &={1 \over 6}V^{-2}C_{IPQ}K^I K^P K^Q+{1 \over 2}
V^{-1}L_I K^I+M\,,  \cr
\nabla \times \omega & = V\nabla M -M\nabla
V+{1 \over 2}(K^I\nabla L_I-L_I\nabla K^I)\,.}}
Indeed, we can check that the solution \zd\ takes this form with
\eqn\ej{\eqalign{ K^I   =  -{q^I \over 2 \Sigma} \,, \qquad
L_I  = 1+{\Qb_I \over 4  \Sigma}\,, \qquad
M ={J_T \over 16}\left({1 \over R}- {1 \over \Sigma}\right)
\,, \qquad  V ={1 \over \rho} \,. }}

The virtue of this is that we can now modify the Gibbons-Hawking
harmonic function to
\eqn\zg{ V = h+{1 \over \rho}}
for constant $h$ and, using \zf\ and \ej, we still have a solution.
Actually, in order to avoid both Dirac string singularities and closed time-like
curves, the relation \jtt\
between $J_T$ and the dipole charges must be modified to:
\eqn\rtnew{J_T  \left(h+{1 \over R}\right) = 4 (q^1 +q^2 +q^3) \,.}
This will be discussed in detail in later sections, but it follows
because the absence of singularities in   $\omega$
puts constraints on the  sources in the third equation of \zf.

For small ring radius, $R\ll 1$, or for small $h$, this reduces to
the five-dimensional black  ring described above.  We now wish to
consider the opposite limit, $R\gg 1$. However, if we want to keep
the quantized charges of the ring fixed, the changing of $R$
should be such that \rtnew\ remains satisfied. We can think of
this as keeping the physical radius of the ring fixed while
changing its position in Taub-NUT. In this limit the black ring is
far from the Taub-NUT tip and effectively sees an infinite
cylinder. In other words, it is physically clear that the 
black ring in this limit becomes a straight black string 
wrapped on a circle, which is
nothing but a four-dimensional black hole.

To see this in more detail, we analyze the geometry in the region
far from the tip, that is, for $\rho \gg 1$, where  we can take
$V=h$.  We also want to center the three-dimensional spherical
coordinates on the ring, and so we change to coordinates such that
$\Sigma$ is the radius away from the ring.  We then have:
\eqn\yb{ d\rho^2 + \rho^2(d\theta^2 +\sin^2 \theta  d\phi^2 ) = d\Sigma^2 +
\Sigma^2 (d\thetah^2 + \sin^2 \thetah d\phi^2)~,} and
\eqn\ya{ \rho = \sqrt{\Sigma^2 + R^2 -2R\Sigma \cos\thetah}, \quad
\cos \theta = {\Sigma \cos \thetah - R \over \rho}~.}

Taking $R \rightarrow \infty$, at fixed $(\Sigma,\thetah,\phih)$ and $h+{1 \over R}$,
we find that the metric is:
\eqn\yy{ds^2= -(\tZ_1\tZ_2\tZ_3)^{-2/3}(d\tt+ \tilde \mu  d\psi)^2
+(\tZ_1 \tZ_2 \tZ_3)^{1/3} \left(dr^2 + r^2(d\thetah^2 + \sin^2
\thetah d\phih^2) \right)\,,}
where
\eqn\coords{  \tZ_I  \equiv {Z_I \over h} \,, \qquad \tilde \mu
\equiv {\mu \over h} \,, \qquad r    \equiv h \Sigma  \,, \qquad
\tt \equiv {t \over h} \,.}
When written in terms of the coordinate $r$ the  harmonic
functions become:
\eqn\yz{ \tZ_I   = {1 \over h} + {\Qb_I \over 4 r} +
{C_{IJK}q^Jq^K \over 8 r^2 } \,, \qquad \tilde \mu    =-{J_T \over
16 r} -{q^I \Qb_I \over 16 r^2} -{q^1 q^2 q^3 \over 8 r^3} \,, \qquad
\omega   =0\,.}
This is precisely the four-dimensional black hole  found by
wrapping the black string solution of \BenaWV\  on a circle.

As noted in \BenaTK, the entropy of the 5D black ring takes a
simple form in terms of the quartic invariant of $E_{7(7)}$, which
gave the first clue to the relation with four-dimensional black
holes, since compactifications to four dimensions have an
$E_{7(7)}$ duality group. The general entropy for this class of
black holes is \KalloshUY:
\eqn\ana{ S = 2\pi \sqrt{J_4}~,}
where $J_4$ is the quartic $E_{7(7)}$ invariant, which can be
expressed in the basis $(x_{ij},y^{ij})$ as
\eqn\ao{\eqalign{ J_4 &= -{1 \over 4}(x_{12}y^{12} + x_{34}y^{34}
+x_{56}y^{56}+x_{78}y^{78})^2-
(x_{12}x_{34}x_{56}x_{78}+y^{12}y^{34}y^{56}y^{78})\cr & +
x_{12}x_{34}y^{12}y^{34}+ x_{12}x_{56}y^{12}y^{56} +
x_{34}x_{56}y^{34}y^{56}+x_{12}x_{78}y^{12}y^{78}+ x_{34}x_{78}
y^{34}y^{78}\cr &+x_{56}x_{78}y^{56}y^{78} .}}
The black-ring entropy is recovered by taking
\eqn\app{\eqalign{x_{12}&=\Qb_1, \quad x_{34}=\Qb_2, \quad
x_{56}=\Qb_3,\ \quad x_{78}=0, \cr  y^{12} &= q^1, \quad
y^{34}=q^2, \quad y^{56} = q^3, \quad y^{78}= J_T = J_{\psit}
-J_{\phit}.}}
Hence, the ``tube angular momentum'' $J_T$ plays the role of
momentum in the four-dimensional black hole picture.  As we
explained above, for the asymptotically-flat 5D black ring, $J_T$
is the difference of the two independent angular momenta, and is
given by \jtt.

We now discuss the implications of this for a microscopic
understanding of the BPS black ring.  In terms of M-theory
compactified on $T^6\times S^1$ the solution \yy\ represents $q^I$
M5-branes wrapped on the $I$'th 4-cycle, and $\Qb_I$ M2-branes
wrapped on the $I$'th 2-cycle.  All the branes are also wrapped on
$S^1$ with momentum $J_T$ flowing on the intersection.  The
microscopic entropy for the most generic type of such a black hole
has been computed in \BertoliniYA\  based on the technology
developed in \MaldacenaDE \foot{It turns out that the analysis of
\MaldacenaDE\ suffices for the particular combination of charges
considered here.}.

Here we have shown that this entropy counting also  applies to the
five-dimensional black ring.  The key point is that the entropy
formula \ana\ is valid throughout the interpolation between the 5D
black ring and the 4D black hole, as follows from the moduli
independence of the entropy.   This yields the same microscopic 
description of black rings as was presented in \BenaTK. 
The difference is that there we had
to rely on taking a near ring limit, which, while physically
plausible, was not as rigorous as the Taub-NUT interpolation given
here.

This analysis also shows that the microscopic
charges and angular momentum of the five-dimensional black ring
are not the same as the charge and angular momenta measured at
infinity. This confirms the observation of \BW\ that the charge
and angular momenta of the black-ring solution have a component
coming from the microscopic charge and microscopic angular
momentum of the ring, and another component coming from charge and
angular momenta carried by the fluxes. This also shows that the
charges defined in \HorowitzJE, while mathematically well defined,
are not the microscopic charges of the black ring.
It would be very interesting to find an interpretation of  the charges
of \HorowitzJE\ in the limit when we are far away from the tip where the ring
becomes a  four-dimensional black hole. One puzzling feature is 
that as one puts together
several four-dimensional black holes, of charges $\Qb_{A,i}$ and $q^A_i$, the
charges defined in \HorowitzJE, given by $Q_{A,i} = \Qb_{A,i} +
{1 \over 2} C_{ABC} q^B_i q^C_i $, are not additive.

A few other issues relating to the entropy deserve comment.  In
\CyrierHJ\ it was proposed to give the black ring a CFT
description based on a four-dimensional black hole CFT with
charges $Q_I$ rather than $\Qb_I$, and with momentum $J_{\psi}$
rather than $J_T$. In order to recover the entropy formula \ana\
an important role was played by a non-extensive zero point energy
shift of $L_0$.  In light of the present understanding it is
rather mysterious to us why this gives the right entropy, since we
have shown explicitly that the relevant four-dimensional black
hole CFT is the one with charges $\Qb_I$, momentum $J_T$, and no
zero point shift of $L_0$. We should also note that the
microscopic description given here trivially carries over to the
case of multi-ring solutions \refs{\GauntlettWH,\GauntlettQY},
yielding the correct entropy. On the other hand, the approach of
\CyrierHJ\  would seem to run into problems since the total charge
$Q_{A}$ is not simply a sum of the individual $Q_{A,i}$, but gets
contributions from cross terms of the form $C_{ABC}~q_i^B~ q_j^C$.

\newsec{Three-charge solutions with Gibbons-Hawking base}

\subsec{The general system of equations}

As shown in \refs{\AdSBH,\BW}, an M-theory background that
preserves the same supersymmetries as three orthogonal M2-branes
can be written as:
\eqn\background{\eqalign{ ds_{11}^2& =  - \left({1 \over Z_1 Z_2
Z_3}\right)^{2/3} (dt+k)^2 + \left( Z_1 Z_2 Z_3\right)^{1/3}
h_{mn}dx^m dx^n \cr &+ \left({Z_2 Z_3 \over
Z_1^2}\right)^{1/3}(dx_1^2+dx_2^2) + \left({Z_1 Z_3 \over
Z_2^2}\right)^{1/3}(dx_3^2+dx_4^2) + \left({Z_1 Z_2 \over
Z_3^2}\right)^{1/3}(dx_5^2+dx_6^2) \,, \cr &   \cr {\cal A} & =
A^1 \wedge dx_1 \wedge dx_2 +A^2 \wedge dx_3 \wedge dx_4 + A^3
\wedge dx_5 \wedge dx_6~,}}
where $A^I$ and $k$ are one-forms in the five-dimensional space
transverse to the $T^6$.  The metric $h_{mn}$ is a
four-dimensional hyper-K\"ahler metric.

When written in terms of the ``dipole field strengths''    $\Theta^I$,
\eqn\Thetadefn{\Theta^I \equiv d A^I + d\Big(  {dt +k \over
Z_I}\Big)\,, }
the BPS equations simplify to \refs{\AdSBH,\BW}:
\eqn\eom{\eqalign{ \Theta^I  &= \star_4 \Theta^I \cr \nabla^2  Z_I
& = {1 \over 2  } C_{IJK} \star_4 (\Theta^J \wedge \Theta^K) \cr
dk + \star_4 dk &= Z_I \Theta^I~,}}
where $\star_4$ is the Hodge dual taken with respect to the
four-dimensional metric $h_{mn}$.
We will take the base to have a Gibbons-Hawking metric:
\eqn\ad{h_{mn}dx^m dx^n  =V(dr^2 + r^2 d\theta^2 + r^2 \sin^2
\theta d\phi^2)+{1 \over V}\big( d\psi + \vec{A} \cdot
d\vec{y}\big)^2}
with
\eqn\AVreln{\vec \nabla \times \vec A ~=~ \vec \nabla V.}
Here we will, for the present, consider a completely general base with
an arbitrary harmonic function, $V$.  We will denote the one-form,
$ \vec{A} \cdot d\vec{y} \equiv A$.

This metric has the following orthonormal basis of 1-forms:
\eqn\sdframes{\hat e^1~=~ V^{-{1\over 2}}\, (d\psi ~+~ A) \,,
\qquad \hat e^{a+1} ~=~ V^{1\over 2}\, dy^a \,, \quad a=1,2,3 \,.}
There are two natural  sets of two-forms:
\eqn\sdforms{\Omega_\pm^{(a)} ~\equiv~ \hat e^1  \wedge \hat
e^{a+1} ~\pm~ \coeff{1}{2}\, \epsilon_{abc}\,\hat e^{b+1}  \wedge
\hat e^{c+1} \,, \qquad a =1,2,3\,.}
The $\Omega_-^{(a)}$ are anti-self-dual and harmonic, defining the
hyper- K\"ahler  structure on the base.   $\Omega_+^{(a)}$ are
self-dual, and we can take the  self-dual field strengths
$\Theta^I$ to be proportional to them:
\eqn\thetaansatz{\Theta^{I} ~=~ - \, \sum_{a=1}^3 \,
\big(\partial_a \big( V^{-1}\, K^I \big)\big) \, \Omega_+^{(a)}
\,.}
For $\Theta^I$ to be closed, the functions $K^I$ have to be
harmonic in $\IR^3$.   Potentials satisfying $\Theta^I = dB^I$ are
then:
\eqn\Thetapots{ B^I ~\equiv~  V^{-1}\,   K^I   \,
(d\psi ~+~ A) ~+~ \vec{\xi}^I \cdot d \vec y \,,}
where
\eqn\xidefn{ \vec  \nabla \times \vec \xi^I ~=~ - \vec \nabla K^I\,.}
Hence, $\vec \xi^{I}$ are vector potentials for magnetic monopoles
located at the poles of $K^I$.

The three self-dual Maxwell fields $\Theta^I$ are thus determined
by the three harmonic functions $K^I$.   Inserting this result in
the right hand side of \eom\ we find:
\eqn\zzres{Z_I ~=~ \half C_{IJK} {K^J K^K \over V} ~+~ L_I \,,}
where $L_I$ are three more independent harmonic functions.

We now write the one-form $k$ as:
\eqn\kansatz{k ~=~ \mu ( d\psi + A   ) ~+~ \omega}
and then the last equation in \eom\ becomes:
\eqn\roteqn{ \vec \nabla \times \vec \omega ~=~  ( V \vec \nabla \mu ~-~
\mu \vec \nabla V ) ~-~ \, V\, \sum_{i=1}^3 \,
 Z_I \, \vec \nabla \bigg({K^I \over V}\bigg) \,.}
Taking the divergence yields the following equation for $\mu$:
\eqn\mueqn{   \nabla^2 \mu ~=~ 2\, V^{-1}\, \vec \nabla \cdot
\bigg( V \sum_{i=1}^3 \, Z_I ~\vec \nabla {K^I \over V} \bigg)
\,,}
which is solved by:
\eqn\mures{\mu ~=~ {1 \over 6} C_{IJK} {K^I K^J K^K \over V^2} ~+~
{1 \over 2 \,V} K^I L_I ~+~  M\,,}
where $M$ is yet another harmonic function.  Indeed, $M$
determines the anti-self-dual part of $dk$ that cancels out of the
last equation in \eom. Substituting this result for $\mu$ into
\roteqn\ we find that $\omega$ satisfies
\eqn\omegeqn{\vec \nabla \times \vec \omega ~=~  V \vec \nabla M -
M \vec \nabla M +  {1\over 2}(K^I  \vec\nabla L_I - L_I \vec
\nabla K^I )\,.}

Summarizing,  to construct a 
solution with three charges, three dipole charges, and a
Gibbons-Hawking base one needs to specify {\it eight}  harmonic
functions in $\IR^3$: $V,K^I,L_I$ and $M$. The function $V$ gives
the Gibbons-Hawking base, the $K^I$ determine the ``dipole''
charges, the $L_I$ represent a contribution to the charges, and
$M$ determines the anti-self-dual part of  $dk$.

The result of our analysis reproduces that of \refs{\min,
\GauntlettQY}.  Here we have proceeded  via the linear algorithm
spelled out in \BW: first computing $\Theta^I$ and then  using
these as  sources in the equations for $Z_I$. This procedure
highlights an important ``gauge invariance'': we can add to  $K^I$
an arbitrary multiple of $V$ without affecting  $\Theta^I$ (which
only depend on the derivatives of $K^I \over V$). Hence, this
change does not affect the solution. At the level of the
equations, such a shift can be reabsorbed by shifting the
 $L_I$ and $M$ by appropriate harmonic functions, and does not
 affect the physical
quantities $Z_I$, $\mu$, $\omega$, and $\Theta^I$. The gauge
invariance of the solutions can be written as:
\eqn\gauge{\eqalign{K^I &~\ra~   K^I ~+~  l^I\, V \,,\cr L_I
&~\ra~ L_I ~-~  C_{IJK}l^J K^K ~-~ \half C_{IJK}l^J l^K V \,,\cr M
&~\ra~ M -\half l^I L_I +{1 \over 12}C_{IJK}\left( V l^I l^J l^K
+3 l^I l^J K^K\right) \,, }}
where the $l^I$ are three arbitrary constants.
This gauge invariance can be used to eliminate one of
the terms that will appear in $K^I$, and greatly simplifies the
solution. The system of equations above gives all solutions with a
Gibbons-Hawking base.

\subsec{Solving for $\omega$}

Since everything is determined by eight harmonic functions, all
that remains is to solve for $\omega$ in equation \omegeqn. The
right hand side has only terms of the form $H_1 \nabla H_2 - H_2
\nabla H_1$, where $H_1$ and $H_2$ are harmonic functions sourced
at discrete points in the base, and possibly have an overall
additive constant.   Let $\vec{y}^{(j)}$ be  the positions of the
source points in the base, and let  $r_j \equiv
|\vec{y}-\vec{y}^{(j)}|$. The right-hand side of \omegeqn\
therefore has terms of the form:
\eqn\rhs{{1\over r_i}\, \vec \nabla\, {1\over r_j} ~-~ {1\over r_j}\, \vec \nabla\,
{1\over r_i}   \qquad {\rm and } \qquad \vec \nabla \, {1\over r_i} \,.}
Hence  $\omega$  will be built from
the vectors $ \vec c_{ij}$ and $\vec v_{i}$ satisfying
\eqn\c{ \vec \nabla \times \vec c_{ij} ~=~  {1\over r_i}\, \vec
\nabla\, {1\over r_j} ~-~ {1\over r_j}\, \vec \nabla\, {1\over
r_i} \qquad {\rm and } \qquad \vec  \nabla \times \vec v_{i} ~=~
\vec \nabla {1\over r_i}~.}

To find $ \vec c_{ij}$ and $\vec v_{i}$ one therefore only needs to
consider pairs of source points, like $(\vec y^{(1)}, \vec y^{(2)})$.
Choose coordinates so that $\vec y^{(1)} ~=~ (0,0,0)$ and $\vec
y^{(2)} ~=~ (0,0,-R)$. Then the explicit solutions may be written
very simply. Let $(y_1,y_2, y_3) = (x,y,z)$, $\Sigma= \sqrt{x^2 +
y^2 +(z+R)^2}$ , and let $\phi$ denote the polar angle in the
$(x,y)$-plane.  Then:
\eqn\vz{ v_{1} ~=~  {z \over r} \, d \phi \,, \qquad  v_{2} ~=~
{(z +R) \over \Sigma } \, d \phi \,,}
and
\eqn\cza{ c_{12} ~=~ - {(r^2 +  R \, z) \over R \, r  \, \Sigma}
\, d \phi \,.}
One then converts these back to a more general system of coordinates
and then adds up all the contributions to $\omega$ from all the
pairs of points.

\subsec{Absence of closed timelike curves}

We have now obtained the general solution, but we
still need to restrict our choice of harmonic functions to avoid
unphysical pathologies.  In particular, we consider the necessary conditions
to avoid the appearance  of closed timelike curves (CTCs).

 If we define $W \equiv (Z_1 Z_2 Z_3)^{1/6}$, and use
the expression for $k$ in \kansatz\ then the metric along the time
and four spatial directions of the base is
\eqn\dsfive{\eqalign{ d s_5  =  & - W^{-4}\big(dt + \mu (d \psi +A) +
\omega \big)^2  +{W^2 V^{-1}} \, ( d\psi + A)^2 \cr
&+ W^2 V \big(dr^2 + r^2 d\theta^2 + r^2 \sin^2
\theta d\phi^2\big)~.}}

To look for CTC's in this metric, it is useful to examine the
spatial components along the Taub-NUT directions:
\eqn\dstil{\eqalign{ d \ts_4   ~=~ & - W^{-4}\, \big( \mu  (d \psi+ A ) +
\omega  \big)^2 \cr & \qquad ~+~ {W^2 V^{-1}}\big(
d\psi + A \big)^2 + W^2 V \big(dr^2 + r^2 d\theta^2 + r^2
\sin^2 \theta \, d\phi^2\big) \cr
 ~=~ & W^{-4}\,  (W^6 V^{-1}-\mu^2 )\Big(  d\psi + A  -
{\mu \,  \omega  \over W^6 V^{-1}-\mu^2 }  \Big)^2  -
 {W^2 \, V^{-1}   \over W^6 V^{-1}-\mu^2 } \,  \omega^2\cr
& ~+~  W^2 V \big(dr^2 + r^2 d\theta^2 + r^2
\sin^2 \theta\, d\phi^2\big) \cr ~=~ & {\CQ \over W^4 V^2} \Big(
d\psi + A  - {\mu \, V^2 \over \CQ }\, \omega \Big)^2 +
W^2 V \Big( r^2 \sin^2 \theta \, d \phi^2 -{\omega^2  \over \CQ} \Big)
+ W^2 V (dr^2 + r^2 d\theta^2) }}
where we have introduced
\eqn\qq{\CQ \equiv W^6 V-\mu^2 V^2 = Z_1 Z_2 Z_3 V-\mu^2 V^2~.}

Hence the closed timelike curves of  the solution are controlled
by two quantities: $\CQ$ and $ r^2 \sin^2 \theta d\phi^2 -{\omega^2
\over \CQ}$, which must thus both be everywhere positive to avoid
CTC's. In fact, upon evaluating $\CQ$ as a function of the eight
harmonic functions that determine the solution we reveal a
beautiful result:
\eqn\qqq{\eqalign{ \CQ &= - M^2\,V^2   - {1 \over
3}\,M\,C_{IJK}{K^I}\,{K^J}\,{K^k} - M\,V\,{K^I}\,{L_I} - {1 \over
4}(K^I L_I)^2\cr &\quad+{1 \over 6} V C^{IJK}L_I L_J L_K +{1 \over
4} C^{IJK}C_{IMN}L_J L_K K^M K^N}}
with $C^{IJK}=C_{IJK}$. We find that  $\CQ$  is nothing other than the
$E_{7(7)}$ quartic invariant \ao\ where the $x$'s are identified
with $L_1,L_2,L_3,-V$, and the $y$'s with $ K_1,K_2,K_3,2M$.

Positivity of $ r^2 \sin^2 \theta \,  d \phi^2 -{\omega^2 \over
\CQ}$ can be examined by looking near the axis between every pairs
of points $y^{(i)}$ and $y^{(j)}$ described above.  
Here we will focus on the two-center solution, and use \vz\ and
\cza.  Positivity implies that $\omega$ should vanish at $\sin
\theta = 0$, that is, on the $z$-axis.  For $R>0$,  the only
combination of  $c_{12}$, $v_1$, and $v_{2}$  that vanishes on the
$z$-axis is:
\eqn\vanish{\eqalign{\omega_0 & ~=~ ( v_{1} + R \, c_{12} -
  v_{2} + d\phi)\cr
&~= ~- {(x^2 + y^2 - (r+z)\, (z+R - \Sigma )) \over r \, \Sigma}\,
 d\phi~.}}
Hence, in order for the solution to have no closed timelike
curves, $\omega$ must be proportional to $\omega_0$.

Another obvious implication of the positivity of $\CQ$ is that  if
we have no charges at the location of the singularities of $V$
({\it i.e.}  $Z_I$ are finite there), then $\mu$ at those locations must
vanish.

\newsec{The General Black Ring in Taub-NUT}

We are now ready to construct the general three-charge,
three-dipole-charge, BPS black ring on  Taub-NUT. These rings are
the generalization of the Taub-NUT supertube of \BenaAY.

To construct a black ring in Taub-NUT we take the harmonic
function $V$ to be sourced at the origin, while allowing the other
harmonic functions to be sourced both at the origin and at the
location of the ring. Some of the harmonic functions will also
contain constants.

We  take the Taub-NUT potential to be:
\eqn\tnbr{ V ~=~  h ~+~ {Q \over r}\,.}
Allowing a general additive constant, $h$,  is useful for
interpolating between the five-dimensional and four-dimensional
descriptions, as we saw in Section 2.

The other harmonic functions can be sourced both at the origin of
Taub-NUT, and at the location of the ring, $\vec y^{(2)} ~=~
(0,0,-R)$. The harmonic functions that give the dipole charges are
thus:
\eqn\brk{K^I ~=~  {2\, \alpha^I \over r} ~+~  {2\, \beta^I \over
\Sigma} ~+~ 2 \,\gamma^I\,.}

To simplify this, we use the gauge invariance \gauge\ to set
$\alpha^I=0$. The $\gamma^I$ have  a  nice interpretation as the
Wilson loops corresponding to  the three gauge fields $A^I$
wrapping the Taub-NUT circle at infinity, and we will keep them
for completeness.   Thus:
\eqn\brksimpl{K^I ~=~  {2 \, \beta^I \over \Sigma} ~+~  2\,
\gamma^I \,.}

By requiring that no charge be present at the origin of the
Taub-NUT space we can set the coefficient of the  $r^{-1}$ term in
the harmonic function $L_I$ to be zero.  Hence
\eqn\Lires{L_I ~=~  {g_I \over \Sigma} ~+~  c_I ~-~ {2  \over
h}C_{IJK}\gamma^J \gamma^K\,.}
The coefficient of the $\Sigma^{-1}$ term contributes to the
charge density of the ring. The constant was chosen such that
$c_I$ are the asymptotic values of the $Z_I$ at infinity, which we
allow to be arbitrary.

Since the $Z_I$ are not sourced at the origin, $\CQ$ can only
be positive if $\mu$ vanishes at $r=0$. This
first implies that the coefficient of the $r^{-1}$ term in the
harmonic function $M$ is zero (otherwise $\mu$ would blow up). The
vanishing of $\mu$ also relates the constant part of $\mu$ and the
coefficient of the $\Sigma^{-1}$ term. If $R>0$, then
\eqn\MM{M ~=~  {K \over \Sigma} ~-~ {K  \over R} \,.}
The coefficient $K$ contributes to the equivalent  of the ``tube''
angular momentum $J_{T}$. The value of $\mu$ at infinity is:
\eqn\muinf{\lim_{r\ra \infty}\mu ~=~ -{K \over R}  ~-~ {2 \over 3
h^2} C_{IJK}\gamma^I \gamma^J \gamma^K ~+~{1 \over h}{c_I
\gamma^I} \,,}
and its effect is to reduce the size of the compactification circle.

We could now begin finding the coefficients of the various terms
appearing on the right hand side of \omegeqn\ and building up
$\omega$ from $v_{1,2}$, and $ c_{12}$ such that \omegeqn\ is
satisfied. We would then have to choose the integration constants
and relations between the parameters of the solutions such that
$\omega$ vanishes on the $z$ axis.

However, it is easier to work backwards, and begin by recalling
that the absence of CTC's implies that $\omega$ must be
proportional to $\omega_0$ defined in \vanish. We then use
equation \omegeqn\ to determine the proportionality coefficient in
terms of the charges of the solutions
\eqn\omegabr{ \omega  ~=~ {K Q \over R}\, \omega_0~,}
and  to find the relation that determines the radius of the black
ring as a function of the charges, dipole charges and moduli:
\eqn\radius{ K \left( h+ {Q \over R} \right) = -  2\,h^{-1}\,
C_{IJK} \beta^I \gamma^J\gamma^K +  c_I \beta^I - g_I \gamma^I\,.}
If we fix the moduli to their value in Section 2
($\gamma_i=0,~Q=1$ and $c_i=1$), and note that in that case:
\eqn\link{\beta^I = -{q^I \over 4}~,~~~ g_I = {\Qb_I \over 4}
~,~~~K = -{J_T \over 16}\,,}
equation \radius\ reproduces the radius formula \rtnew.

One can also compute the value of the entropy for the most
generic choice of moduli, and obtain
\eqn\smod{S = 2 \pi \sqrt{64 \cal J} }
where 
\eqn\cphi{\eqalign{ {\cal J} & ~=~-(\beta^I \, g_I)^2 +
C^{IJK} \, C_{IMN}\, g_J \, g_K \, \beta^M \, \beta^N  ~-~ {8\over 3}  K\, C_{IJK}
\,\beta^I\,\beta^J\, \beta^K \cr 
&~=~- ~\big(g_1 \, \beta^1 + g_2 \,\beta^2 + g_3\, \beta^3 \big)^2
~-~ 16 K \,\beta^1\,\beta^2\, \beta^3 \cr
&~~~~~\, +~ 4 g_1 \, \beta^1 \, g_2 \, \beta^2 
~+~ 4 g_1 \, \beta^1 \, g_3\, \beta^3 
~+~ 4 g_2\,  \beta^2 \, g_3\, \beta^3 \,.}}
The relation between $\beta^I$ and the dipole charges cannot
depend upon the moduli because they are determined by the periods
of the field strengths taken over the two-cycles.  Hence, if the
relation between the M2 charges, $\Qb_I$, and the parameters
$g_I$, and between $J_T$ and $K$, remains the one in equation
\link\ for arbitrary moduli, then one can explain the entropy
\smod\ in exactly the same way as before. However, it is also
possible, although unlikely, that when the $\gamma^I$ are non-zero
the $\Qb_I$ and $J_T$ could be given by some other  more
complicated combinations of the parameters, such that the
$E_{7(7)}$ form of \cphi\ is preserved when  written in terms of
the $\Qb_I$ and $J_T$. We leave the exploration of this and other
issues to future work. Here we simply note that for the solutions
that reduce to five-dimensional black ring solutions in the near-tip
limit (like the one considered in section 2), 
all the $\gamma^I$ vanish, and no such ambiguity exists.

\newsec{Conclusions and Future Directions}

We have presented the solution for a general black ring in
Taub-NUT.  The general solution carries six charges, angular
momentum, and has three moduli corresponding to the sizes of 
two-cycles in $T^6$ and the Taub-NUT fibre, and three moduli corresponding 
to Wilson lines along the fibre. 
We have also identified a gauge invariance in the equations
governing these rings, which can be used to show that our solution is
the most general solution describing a BPS black ring in Taub-NUT.

When the ring is localized near the origin of
Taub-NUT the solution reduces to the five-dimensional BPS black ring, while in the
opposite limit,  where the ring is far from the origin, we recover the four-dimensional
black hole corresponding to a black string wrapped on a circle.

By being able to smoothly interpolate between these two
descriptions we resolved some confusion regarding the microscopic
description of black rings.  In particular, in \BenaTK\ the black
ring entropy was accounted for in the $(4,0)$  CFT of a four-dimensional black
hole with charges $\Qb_I$.  For the five-dimensional black rings $\Qb_I$ are the
``local'' charges of the ring, and differ from the conserved
charges $Q_I$ measured at infinity. However, these charges are hard to define
as flux integrals, and this led to some debate regarding the validity of the
description of $\Qb_I$ as a ring charge.
Now this issue can be resolved unambiguously: by taking the black
ring out to large radius in the Taub-NUT space we see explicitly
that it turns into a four-dimensional black hole with charges $\Qb_I$, and this
makes it clear that the CFT description should be based on these
charges. Furthermore, since the charges are quantized this
conclusion continues to hold even when we bring the ring back near
the origin, where it becomes effectively five-dimensional.

It is straightforward to generalize our solution to the case of
multiple rings, along the lines of \refs{\GauntlettWH,\GauntlettQY}.  
One simply adds
additional source points in the harmonic functions.  The CFT
description is just that of a collection of four-dimensional black hole CFT's,
each with the corresponding charge $\Qb_I$.  The key point is
that it is the $\Qb_I$ and the $q^I$ which {\it are additive}, 
while the
$Q_I = \Qb_I + \half C_{IJK}q^J q^K$ {\it are not}.  This, together with the
fact that the entropy of separated rings is additive implies
that any putative CFT
description based on the $Q_I$ will need to have a mechanism for
disentangling the contributions of each individual ring, which
seems rather difficult to achieve.

As a final comment we should note that in the type IIB frame the
black rings have a near horizon limit which is asymptotically
AdS$_3 \times S^3$. Therefore, the black rings exist as states in
the same  D1-D5 CFT as the usual five-dimensional black hole. 
A microscopic description of the black rings
in this CFT was proposed in \BenaTK. However, this proposal 
rested on one phenomenological
assumption regarding the length of effective strings in the CFT.
Perhaps this assumption can be established more firmly using the
ideas developed here.

\bigskip
\noindent {\bf Acknowledgments:} \medskip \noindent
The work of IB
and PK is supported in part by the NSF grant  PHY-00-99590. The work of NW
is supported in part by the DOE grant DE-FG03-84ER-40168.

\listrefs
 \end